# Distance-based Analysis of Machine Learning Prediction Reliability for Datasets in Materials Science and Other Fields

Evan Askanazi and Ilya Grinberg

Department of Chemistry, Bar Ilan University, Ramat Gan, Israel

**Abstract**

Despite successful use in a wide variety of disciplines for data analysis and prediction, machine learning (ML) methods suffer from a lack of understanding of the reliability of predictions due to the lack of transparency and black-box nature of ML models. In materials science and other fields, typical ML model results include a significant number of low-quality predictions. This problem is known to be particularly acute for target systems which differ significantly from the data used for ML model training. However, to date, a general method for characterization of the difference between the predicted and training system has not been available. Here, we show that a simple metric based on Euclidean feature space distance and sampling density allows effective separation of the accurately predicted data points from data points with poor prediction accuracy. We show that the metric effectiveness is enhanced by the decorrelation of the features using Gram-Schmidt orthogonalization. To demonstrate the generality of the method, we apply it to support vector regression models for various small data sets in materials science and other fields. Our method is computationally simple, can be used with any ML learning method and enables analysis of the sources of the ML prediction errors. Therefore, it is suitable for use as a standard technique for the estimation of ML prediction reliability for small data sets and as a tool for data set design.

**Introduction**

The use of machine learning has become increasingly popular in material science due to the capability of machine learning methods to capture main trends in the data set by fitting complex nonlinear models. Machine learning methods have similarly been applied in a wide variety of other disciplines for data analysis and prediction. While deep learning methods have seen rapid development in the past decade, they cannot be applied for relatively small data sets that are common in materials science and other fields. For such data sets, traditional ML methods such as support vector regression, random forest and XGboost [1-3] must be used. Despite their success, ML methods have several disadvantages such lack of interpretability and inability to accurately estimate the reliability of ML predictions, in contrast to methods based on fundamental scientific principles (e.g. quantum mechanics) where accuracy of prediction can be estimated based on the approximations of the method [4]. For example, for a data set of formation energies of transparent conductor oxides materials used in a recent ML study [5], while the ML predictions are quite accurate in most cases, they still show some error in ~10% of the cases and show severe failures in about 2% of the cases (Fig. 1). To estimate the reliability of the prediction for a given system, an ensemble of prediction models can be used to evaluate the standard deviation of the predicted values. However, this method is not always accurate and may not predict the failure of the model for data that is out of the distribution of the training data.

Prediction ability of machine learning invariably depends on the feature space encompassed by the input variables. For numerous data sets, certain machine learning models will predict far more accurately than others in a given variable subspace and vice versa. [6]. In some cases, variable feature subspaces exist where machine learning itself is inherently limited because the relationships between the input and target variables are especially weak in that subspace. Designing a method to identify these subspaces is a critical step in efforts to be able to discover and successfully predict new materials based on the data from the previously identified materials and in the efforts to improve the general reliability of ML methods. In quantitative structure-activity (QSAR) models, this principle is referred to as the domain of applicability [7]. This refers to the ability of QSAR models to identify and predict new compounds based on their shared physical traits with the chemicals which comprise them. Common algorithms for identifying domains of applicability are based on projection matrices formed from molecular descriptors and target compound values. For these types of models, internal validation, from cross validation and bootstrapping, and external validation are considered necessary to demonstrate their reliability [8].

Intuitively, the similarity of the data point to be predicted to the data points used in the training data set should be related to the reliability. Thus, the distance in feature space between the data points should provide information regarding the reliability of prediction. However, the application of this intuitive concept in practice is not straightforward. For example, it is unclear how the similarity metric should be defined. For example, one can imagine that a predicted data point located a moderate distance away from several training data points will be predicted more reliably than a data point that has only one close neighbor and many distant neighbor training points. Thus, a method for properly weighting the effects of different data points located at different distances must be identified. Furthermore, since the features have different weights and often are strongly correlated with each other, it is also unclear what is the proper method for evaluating the distance. In previous QSAR studies, it was shown that that pointwise error is related to the number of neighbor data points inside a certain distance cutoff [9]; however, the cutoff is not known a priori. Another QSAR approach defined an average distance to the data points in the training set [10]. However, this approach may not provide the correct weighting of the close and distant training data points on prediction reliability. A very recent work focusing on ecology modeling suggesting using the distance between the predicted point and its nearest neighbor in the training data set, demonstrating good error predictability [11]. While numerous studies have been performed probing the relationships between predictability and domains of applicability, feature space and pointwise distance [12-17] in some cases using complex models, a general method for predicting the errors of ML models based on distance in feature space is still unavailable.

In this work, we seek to identify a method to gauge the predictive ability of ML models based on simple and general relationships between data points across feature space. Since ML models are essentially tools for interpolation, it is intuitive that data points in feature space that have a large number of known data in their vicinity will be predicted more accurately, while data points in feature space that have few or no known data in their vicinity will be predicted poorly. Therefore, we develop a method for estimating the average sampling density by the known data points of the feature space relevant to the predicted data point. The inverse of the average sampling density is then defined as the average sampling distance. Using 12 different data sets taken from previous applications of ML in materials science and other disciplines, we show that the ML prediction error systematically increases with increasing average sampling distance (lower sampling density), allowing the classification of the predicted data as high-, medium- and low-reliability based on the evaluated average sampling density or sampling distance. This analysis also allows us to identify data points in the data set for which the features used in the data sets are insufficiently descriptive, such that the prediction error is due to poor features rather than to the insufficiently flexible model. The proposed analysis is simple and computationally inexpensive, and can be applied to a wide variety of datasets and ML methods used in various disciplines. It may also serve as a basis for further development and application for reliability analysis of deep learning predictions, for future improvement of ML prediction accuracy through machine learning targeted to specific regions of feature space, and as a tool for data set design.

**Methods and Discussion**

We follow the approach of using the distances in feature space between the data point to be predicted (for which the features are known but the target property is not) and the known data points (for which both the features and the target property values are known). In order to determine the strength of the relationship between machine learning prediction ability and feature space distances, we first need a reliable determination of a data point's location within the feature space. This entails a coordinate transformation of feature space such that the distances between any two data points in the space can be found in the same manner one would for instance in a Cartesian, cylindrical or polar coordinate system. However, for these coordinate systems, the vectors of the various coordinates are orthogonal, making distance evaluation straightforward, for example using the standard Euclidean distance formula for the Cartesian space. However, the features used as different dimension coordinates in machine learning are almost always correlated to some degree and can be highly correlated, making the standard Euclidean formula inaccurate for the evaluation of the true distance. As illustrated in Fig. 2, for a system with highly correlated features, some distances evaluated by the Euclidean formula to be large are in fact small. Therefore, it is necessary to decorrelate or orthogonalize the features to make the standard Euclidean formula applicable.

To orthogonalize the features, we apply the well-known Gram-Schmidt procedure; by applying this technique on the input data features, we obtain an orthonormal basis from the features from which we can obtain an interpretable metric for determining the data points' distance from each other in feature space. Furthermore, it is intuitively obvious that features that are unimportant (i.e. not correlated with the target predicted property) should not influence prediction accuracy and therefore should not contribute to the distance metric, as demonstrated in a recent deep learning study [18]. Therefore, in the distance metric, the features should be weighed by their relative importance. The extent to which the Gram-Schmidt procedure and the weighting by feature importance improve the utility of the distance metric depends on the extent to which the dataset contains features that strongly correlated with each other and/or dominant in relating the input data to the target variable. Thus, we evaluate the distance in feature space with $n$ features $x_k$ ($k=1,n$) between data points $i$ and $j$ as

(1) $d_{ij} = \sqrt{\sum_{j=1}^{n} w_{k'}(x_i - x_j)^2}$

where $x_{k'}$ are the features constructed from features $x_k$ using Gram-Schmidt orthogonalization and $w_{k'}$ are the weights of these features obtained for the ML model trained using $x_{k'}$.

We now consider the question of how to combine the different distances $d_{ij}$ into a single distance metric. We first consider a simple case of a data point $i$ surrounded by $N$ equally spaced points $j$ such that all $d_{ij}$ are the same and are equal to $d$. This is illustrated in Fig. 2b for $N=4$ for the points on the left. In this case, the sampling density of the target property (function) in the vicinity of point $i$ ($S_i$) is

(2) $S_i = N/d.$

Then, if we consider a data point $i$ surrounded by $N$ points $j$ with different distances (Fig. 2b, points on the right, $N=4$), the average sampling density of the target property (function) in the vicinity of point $i$ ($S_i$) will be given by

(3) $S_i = \sum_{j=1}^{N} 1/d_{ij}$

We can therefore define the distance metric $D_i$ (the average sampling distance) that measures the quality or density of the sampling in the vicinity of point $i$ for a dataset with $N$ data points with known target property values as

(4) $D_i = 1/S_i = 1/(\sum_{j=1}^{N} 1/d_{ij})$ .

We expect that larger error will tend to be obtained for predicted data points with larger $D_i$ values. While even for large values of $D_i$ (small sampling density around data point $i$), some predictions will be accurate simply by chance, the errors will be distributed over a larger range, so that a larger MAE will be obtained. By contrast, for small values of $D_i$, the error distributions should be very narrow and MAE values will be small.

We examine this hypothesis for 12 data sets used in previous ML studies in materials science and other disciplines. The seven materials science datasets are the formation energy [5] and band gap energy [5] of TCOs, the activation energy for dilute solutes in crystals [19], the reduced glass transition temperature for alloys [20], the formation energy and band gap of perovskites[21], and perovskite stability [22]. The five data sets from other disciplines are daily bike count [23], frequency variation of Parkinson's patients [24] game actions [25], energy use [26] and forest fire area [27]. For each of these data sets, we divide the data set into training and test sets with 10-fold cross-validation. We then orthogonalize the features of the training set using the Gram-Schmidt procedure to obtain orthogonal features $x_{k'}$ and then use these features to train an SVR model as described in the methods. Next, we predict the target properties $y_i$ for the test set using the trained SVR models and evaluate the errors of the SVR predictions. Then, we evaluate distance metrics $D_i$ for each of the test points and classify all test set data points based on their distances into groups with equal numbers of data points. Finally, we evaluate the MAE of ML prediction for each group and plot the obtained MAE values versus the group number in Figure 3. To demonstrate the effect of the Gram-Schmidt feature orthogonalization and feature weighting on this analysis, we also present the results of MAE of SVR prediction versus $D_i$ evaluated using the original non-weighted and non-orthogonal features. The MAE results of SVR prediction versus $D_i$ evaluated using non-orthogonal but weighted features and orthogonal but non-weighted features are shown in the SI together with the results obtained with non-weighted and non-orthogonal features and the results obtained with weighted and orthogonal features. Examination of the results presented in Figure 3 shows that with the exception of the forest fire data set (Figure 3l), in all cases there is a clear trend of MAE increasing with increasing group number. For about half the datasets (Figs. 3a,e,f,g,i) the trend is quite smooth while for the other half there are strong fluctuations in the MAE superimposed on the overall trend.

To isolate the effect of the orthogonalization and feature weighting from that of the distance metric formula, we compare to the plots of MAE vs $D_i$ calculated without GS orthogonalization and feature weighting. It is observed that for five datasets (Figs. 3a,b,e,g,l) the benefit of GS orthogonalization and feature weighting is either small or non-existent, with essentially equally smooth trends obtained for both $d_i$ evaluation methods. However, for seven data sets (Figs. 3c,d,f,h,i,j,k), GS orthogonalization and feature weighting clearly improves the smoothness of the trend and in particular, for four data sets (Figs. 3f,h,i,j) the use of GS orthogonalization and feature weighting changes very weak or no correlation of

MAE with $D_i$ into a strong dependence. Thus, for these datasets orthogonalization is a crucial step to obtain the correct distance evaluation. This suggests that previous attempts to use distance metrics for reliability characterization that found no or weak relationships between distances in feature space and prediction accuracy may have been affected by feature space distance evaluation that did not take into account the correlations between the features. Comparison of the plots of MAE vs $D_i$ evaluated using GS orthogonalization and feature weighting, GS orthogonalization without feature weighting, feature weighting without GS orthogonalization and with no GS orthogonalization and feature weighting (see SI) shows that GS orthogonalization has the main effect of improvement of MAE vs $D_i$ trends while feature weighting has a minor effect.

Even with the use of GS orthogonalization and feature weighting, for the forest fire data set, no trend with $D_i$ is found. This is most likely due to the poor features used in this data set that lead to larger errors of SVR prediction. For data sets where the features are not well-correlated with the target properties, the suggested analysis procedure will not be useful because the error is not controlled by the sampling density but rather by the fact that the target property is controlled by hidden features that are omitted in the dataset. In that case, even a high sampling density of the known features will not lead to accurate prediction. Thus, our analysis can identify datasets for which better features are necessary. Furthermore, for datasets for which predictions are accurate overall but some outliers with poor accuracy are obtained, our method can identify whether the poor accuracy of data point *i* is due to weak sampling (as indicated by high $D_i$) of the region of the feature space in which the outlier is located or due to the importance of hidden features for this data point as indicated by high density of sampling and low $D_i$.

To demonstrate the effectiveness of our analysis method in more detail, we present the plots of error vs $D_i$ for individual data points (Fig. 4) and plots of predicted versus actual values (Fig. 5) with different groups shown in different colors for the TCO formation and band gap energy datasets from [10] and bike sharing [13] and game action [16] datasets. It is clear that the data in groups 1-2 corresponding to small distances $D_i$ are predicted accurately, with the data points for these groups shown in red falling on the y=x line in the plots in Fig. 4. With increasing group number corresponding to increasing sampling distance and decreasing sampling density, increased deviation from the y=x line is observed with all outliers corresponding to groups 9-10. We note that the ratios between the distances for groups 1-2 and groups 9-10 is between 2 and 3. Thus, even a relatively small decrease in the sampling density has a strong impact on prediction accuracy.

**Discussion**

We now discuss the implications of our results for small-data machine learning. Our results provide a clear confirmation that ML methods are generally not suitable for extrapolating to regions in feature space where little data are available as can be seen by the prevalence of large errors for group 9-10 data. Our method also provides guidance for targeted exploration of feature space experimentally or computationally by identifying regions of feature space with low sampling density. These undersampled regions should be examined in order to obtained improved data sets, whereas obtaining additional data in the high-density sampling region will not be beneficial for improving the prediction accuracy.

Furthermore, our method enables investigation of the sources of ML prediction error and reliability. A given ML model has three sources of errors, namely insufficient model flexibility, insufficiently descriptive features and insufficient sampling of the features space by the data set used for model training. Previously, it was difficult to deconvolute the effects of these error sources. Using our method, we can separate the effects of the sampling density, such that for low $D_i$ there is little sampling error and for high $D_i$, the error is clearly dominated by insufficient sampling. If a high error is found for data points in regions of high sampling (low $D_i$), this indicates insufficiently suitable features or insufficiently flexible model. To distinguish these two error sources, it may be possible to use a series of different ML methods. If a clear improvement is obtained for one method compared to the other, this indicates that model flexibility is important. If all methods obtain the same results, this suggests that model flexibility is not the limiting feature in achieving the desired accuracy or prediction and that rather feature selection must be improved to improve prediction accuracy.

In future work, our local sampling density analysis method may serve as a basis for addressing several other problems in ML learning. First, the method is quite simple and yet it can predict the overall trend error (reliability) of ML prediction for a variety of data sets. It is likely that more sophisticated analyses of the feature space and the dependence of the target properties on the features will provide a more granular and accurate estimate of prediction reliability (e.g. explaining why for groups 1-2 in the bike sharing [23] and game action [25] datasets, some of the data points are predicted with very low error while other are predicted with moderate error). For example, a distance metric function can be suggested that uses a weighting scheme that is different from that expressed by Eq. 4. In this regard, we have examined the use of an exponential weighting function but did not see any improvement (see SI). Nevertheless, more sophisticated approaches may result in improved separation ability.

Additionally, it may be possible to develop more accurate methods for ML prediction by designing separate methods for application to the high-sampling regions that focus on the accurate interpolation of the highly-sampled data and for application to the low-sampling feature space regions that focus on accurately capturing the overall trends. Finally, our method relies on the orthogonalization of the features using the GS procedure which scales as $N^3$ where N is the number of data points. Due to its $N^3$ scaling, GS orthogonalization cannot be applied to big data problems and therefore cannot be used to evaluate the accuracy of deep learning predictions. Therefore, another future direction is to investigate how the distance metric $D_i$ can be evaluated efficiently for big data sets, and whether such distance-metric-based analysis of reliability is useful for problems addressed by deep learning methods.

**Conclusion**

We have demonstrated that the errors of ML prediction for small data sets generally show a well-defined and systematic dependence on their separation in feature space from other data points. For various data sets and multiple independently generated data points, we find that ML prediction error tends to increase with increasing distance metric defined as the inverse of the average local sampling density. We also find that the use of decorrelated features created by the application of the Gram-Schmidt orthogonalization procedure to the features of the data used in ML model strongly increase the accuracy of ML reliability prediction based on the distance metric for some datasets. Our method is computationally simple, can be used with any ML learning method and enables analysis of the sources of the ML prediction errors. Therefore, it is suitable for use as a standard technique for the estimation of ML prediction reliability and design of improved datasets for ML.

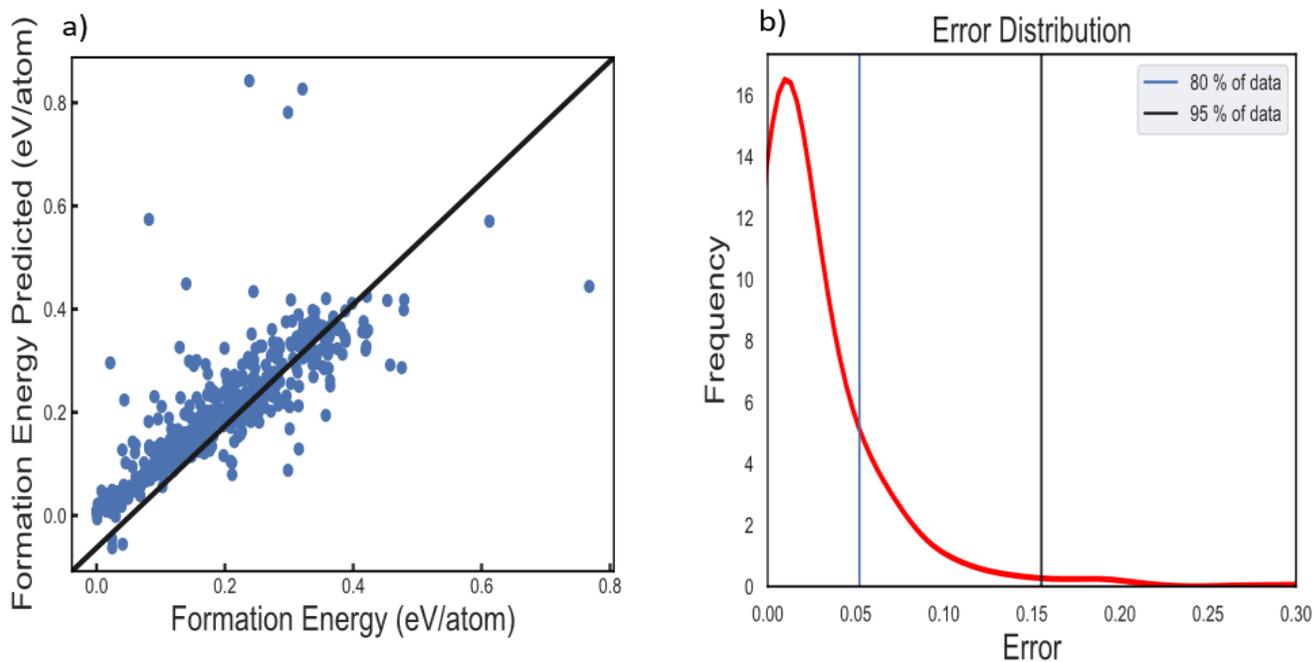

**Figure 1. Reliability of predictions by standard methods.** a) Predicted versus actual formation energies for the TCO formation dataset b) Distribution of the errors in the predicted formation energy values by SVR

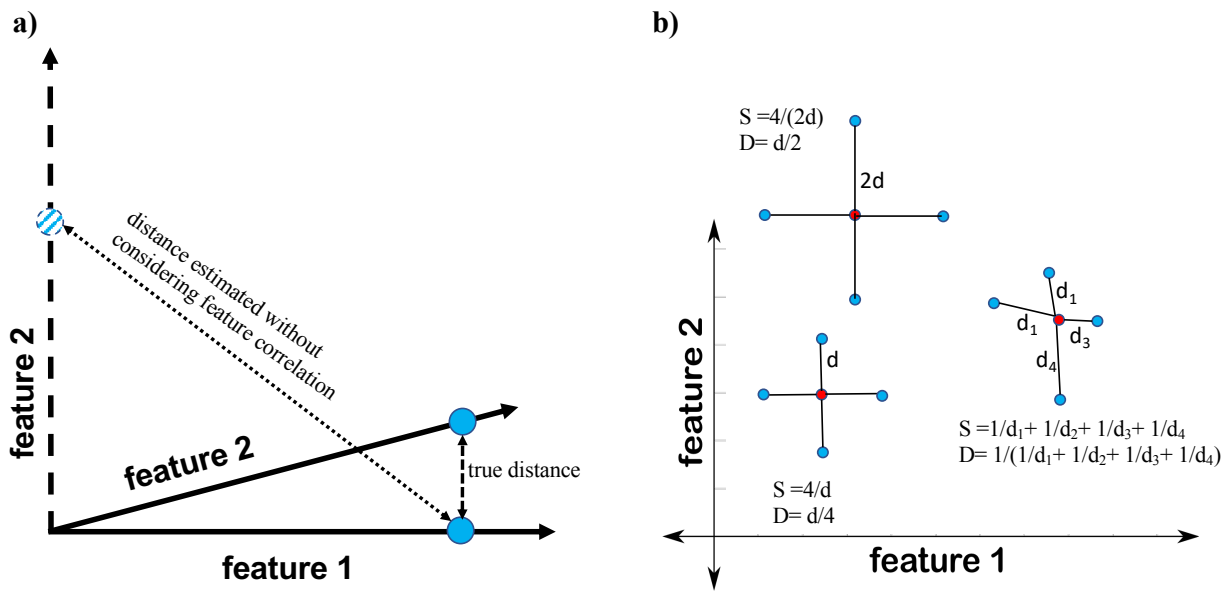

**Figure 2. The underlying logic of the distance metric formula.** a) Illustration of the difference between true distance and distance estimated by simple Euclidean formula for non-orthogonal coordinate systems. b) Illustration of the derivation of the distance metric formula from the local sampling density.

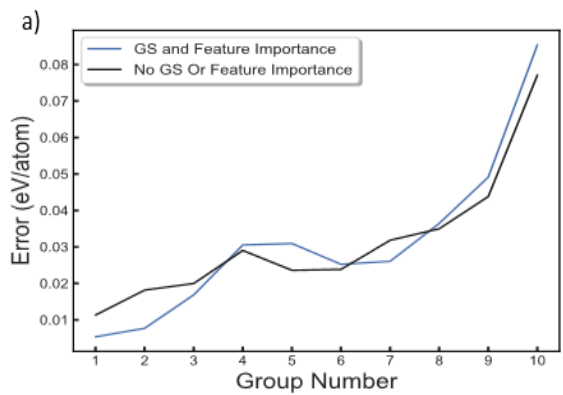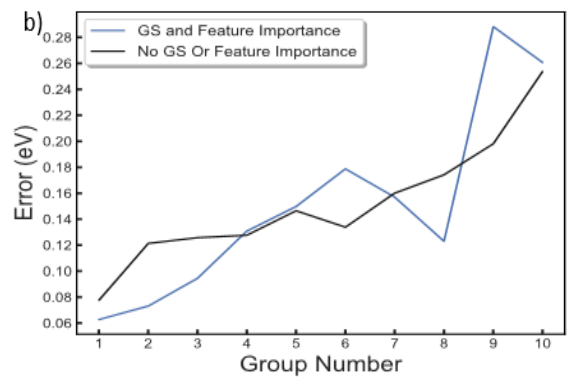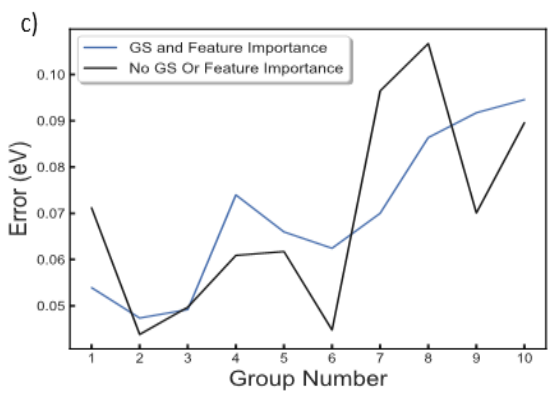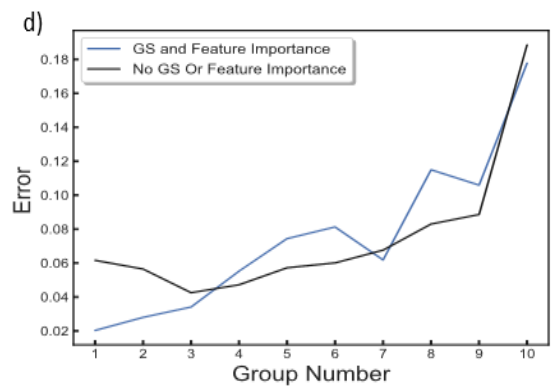

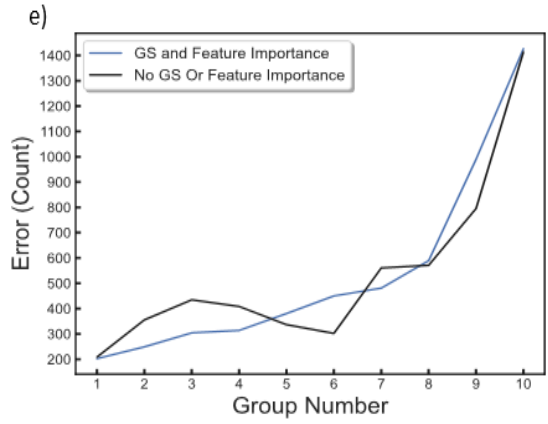
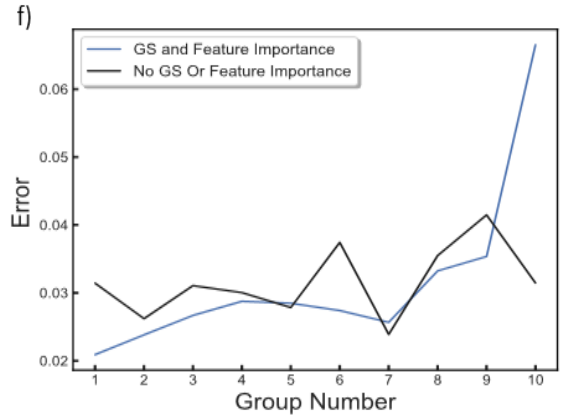
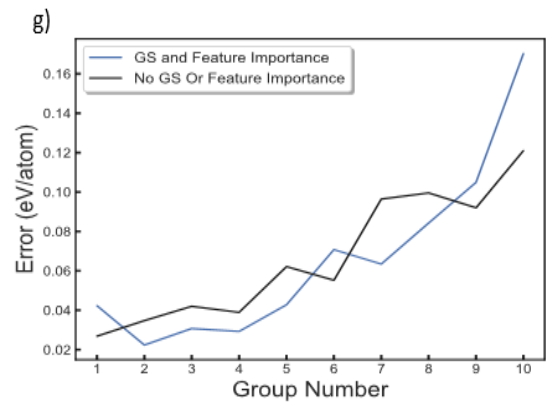
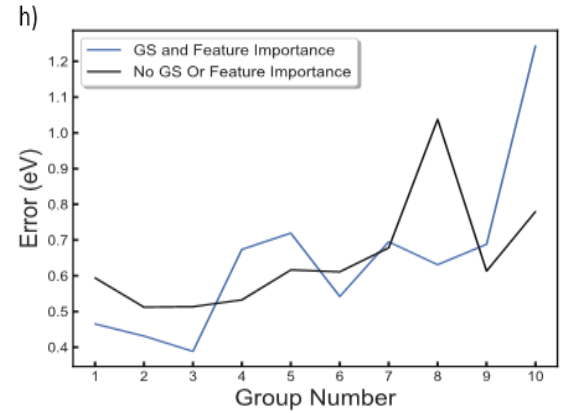

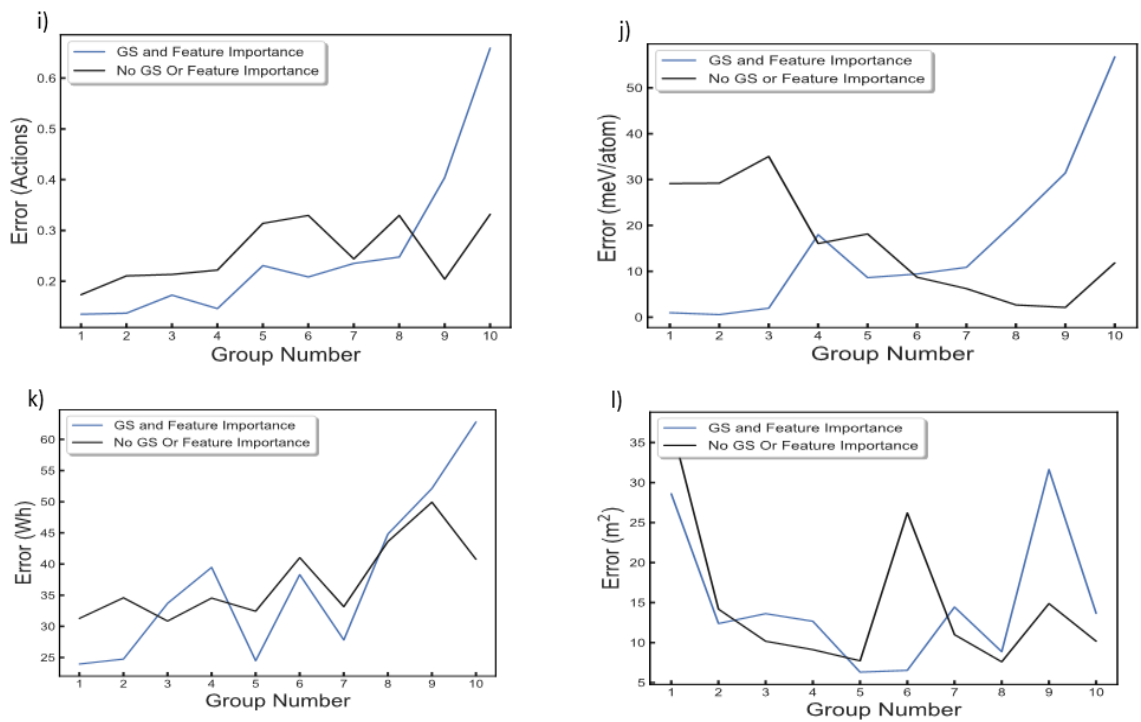

**FIGURE 3. Relationship between the distance metric and prediction error.** Mean absolute prediction errors as a function of distances of data points from all others in orthonormalized feature space. The distances are grouped into 10 subgropus with error standard deviations computed for each subgropus. The computations are done for a) TCO formation energy [5] b) TCO band gap energy [5] c) activation energy for dilute solute in crystal [19] d) reduced glass transition temperature for alloys [20] e) daily bike count [21] f) frequency variation of Parkinson's patients [22] g) perovskite formation energy [23] h) perovskite band gap energy [23] i) game actions [24] j) perovskite stability [25] k) energy use [26] and l) forest fire area [27] datasets.

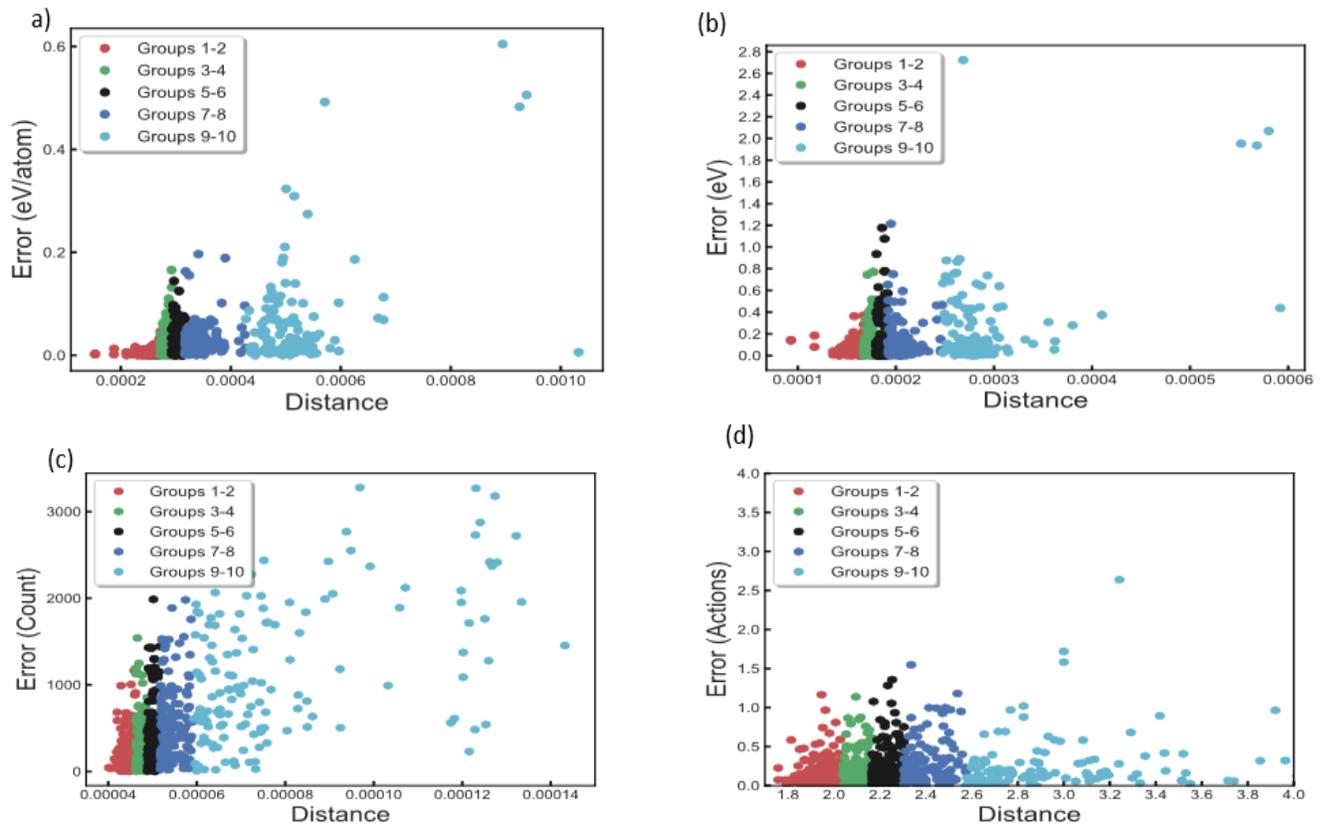

**FIGURE 4. Error distributions as a function of distance.** Errors of individual ML prediction plotted as a function of the distance metric D for the (a) TCO formation energy (b) TCO band gap energy (c) daily bike count and (d) game action datasets with data points for different groups shown by different colors.

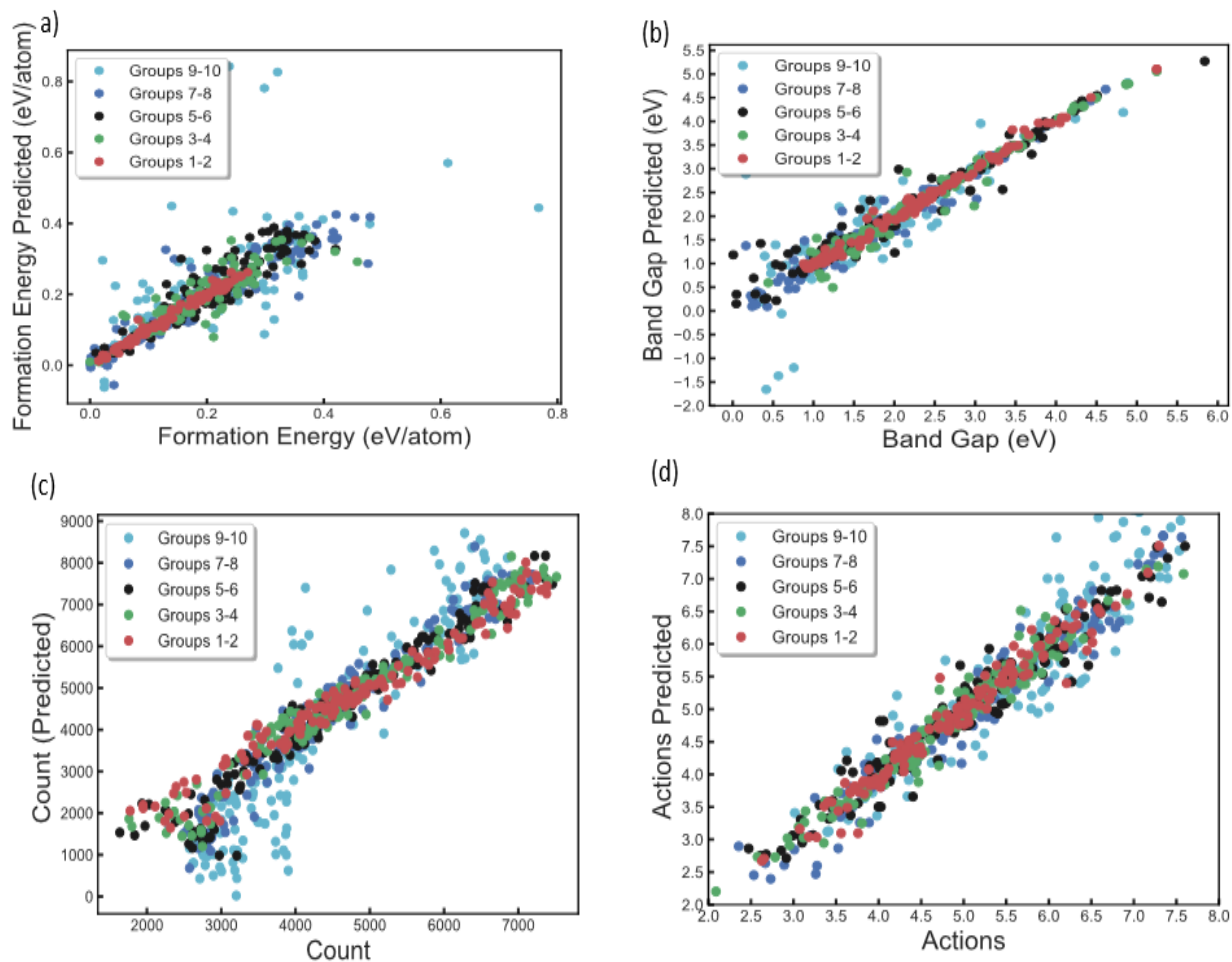

**FIGURE 5. Prediction accuracy for different distances.** Real versus ML-predicted values for the (a) TCO formation energy (b) TCO band gap energy (c) daily bike count and (d) game action data sets with data points for different groups shown by different colors.